\def\papertitle{WaveNet-Style Guitar Amplifier Model Pruning for Real-Time iOS Deployment}
\def\paperauthorA{Ryota Sato}
\def\paperauthorB{Eli Silverstein}

\documentclass[twoside,a4paper]{article}
\usepackage{etoolbox}

\usepackage{dafx26demo}

\usepackage{amsmath,amssymb,amsfonts,amsthm}
\usepackage{siunitx}
\usepackage{euscript}
\usepackage[T1]{fontenc}
\usepackage[utf8]{inputenc}
\usepackage{ifpdf}
\usepackage[english]{babel}
\usepackage{caption}
\usepackage{subfig} 
\usepackage{color}
\usepackage{booktabs}

\input glyphtounicode
\ninept

\newcounter{numauth}
\newcounter{listcnt}
\newcommand\authcnt[1]{\ifdefined#1 \stepcounter{numauth} \fi}

\newcommand\addauth[1]{
\ifdefined#1
\stepcounter{listcnt}
\ifnum \value{listcnt}<\value{numauth}
\appto\authorslist{, #1}
\else
\appto\authorslist{~and~#1}
\fi
\fi}
\authcnt{\paperauthorB}
\authcnt{\paperauthorC}
\authcnt{\paperauthorD}
\authcnt{\paperauthorE}
\authcnt{\paperauthorF}
\authcnt{\paperauthorG}
\authcnt{\paperauthorH}
\authcnt{\paperauthorI}
\authcnt{\paperauthorJ}
\def\authorslist{\paperauthorA}
\addauth{\paperauthorB}
\addauth{\paperauthorC}
\addauth{\paperauthorD}
\addauth{\paperauthorE}
\addauth{\paperauthorF}
\addauth{\paperauthorG}
\addauth{\paperauthorH}
\addauth{\paperauthorI}
\addauth{\paperauthorJ}

\usepackage{times}

\usepackage[final]{microtype}
\usepackage{xurl}
\newif\ifpdf
\ifx\pdfoutput\relax
\else
   \ifcase\pdfoutput
      \pdffalse
   \else
      \pdftrue
   \fi
\fi

\ifpdf 
  \usepackage[pdftex,
    pdftitle={\papertitle},
    pdfauthor={\authorslist},
    pdfsubject={Proceedings of the 29th International Conference on Digital Audio Effects (DAFx26)},
    colorlinks=false, 
    bookmarksnumbered, 
    pdfstartview=XYZ 
  ]{hyperref}
  \usepackage[pdftex]{graphicx}
\else 
  \usepackage[dvips]{epsfig,graphicx}
  \usepackage[dvips,
    pdftitle={\papertitle},
    pdfauthor={\authorslist},
    pdfsubject={Proceedings of the 29th International Conference on Digital Audio Effects (DAFx26)},
    colorlinks=false, 
    bookmarksnumbered, 
    pdfstartview=XYZ 
  ]{hyperref}
\fi
\usepackage[hypcap=true]{caption}
\title{\papertitle}

\affiliation
{\paperauthorA\ and \paperauthorB}
{\href{https://ee.stanford.edu/}{Dept. of Electrical Engineering} \\ Stanford University \\ Stanford, USA\\
{\tt \href{mailto:ryos17@stanford.edu}{ryos17@stanford.edu}} \\
{\tt \href{mailto:esilvers@stanford.edu}{esilvers@stanford.edu}}
}

\begin{document}
\ifpdf 
  \DeclareGraphicsExtensions{.png,.jpg,.pdf}
\else  
  \DeclareGraphicsExtensions{.eps}
\fi

\maketitle

\begin{abstract}
WaveNet-style convolutional networks emulate tube amplifiers and distortion pedals with high fidelity, but their computational cost has confined them to desktops or dedicated DSP hardware. We present a sparse-enabled WaveNet inference engine for iOS that runs heavily pruned neural guitar amplifier models in real time on iPhones. Aggressive iterative magnitude pruning removes 90\% of the network weights with no perceptible loss in quality. A custom sparse C++ engine turns this sparsity directly into compute savings, sustaining low-latency real-time operation on a CPU-only iPhone implementation where the dense model cannot. On-device output matches the trained model to within int16 quantization error. At the demonstration, visitors will play a guitar through the app on iPhone hardware and A/B the on-device pruned model against the physical pedal it emulates. Source code and audio examples are available at \url{https://github.com/ryos17/wavenet-imp}.
\end{abstract}

\section{Introduction}
\label{sec:intro}

Virtual analog (VA) modeling of audio circuits has become an active research area, particularly for guitar amplifiers and effects~\cite{virtualanalogeffects}. White-box methods explicitly simulate the underlying circuit~\cite{wavesim} but require detailed component knowledge, whereas black-box methods learn the nonlinear input--output relationship directly from measured data. Neural networks are especially effective here, with early approaches using Long Short-Term Memory (LSTM) recurrent networks~\cite{lstm-1, lstm-2} and state-of-the-art methods using WaveNet-style convolutional networks with dilated one-dimensional convolutions~\cite{guitar-wavenet}.

Such models have traditionally been deployed on desktop machines or dedicated hardware~\cite{quad_cortex} because of their high computational cost. To improve efficiency, prior work has turned to iterative magnitude pruning~\cite{iterative-magnitude-pruning}, which removes a large fraction of weights while preserving accuracy, as shown for LSTM-based neural amp models~\cite{pruning-lstm}. Such sparse subnetworks are supported by the lottery ticket hypothesis~\cite{lottery-ticket} and confirmed for convolutional audio models by Esling et al.~\cite{esling-lottery}. However, systematic pruning of WaveNet-style neural amplifiers remains largely unexplored, and little effort has targeted sparsity-aware inference engines for neural amplifier modeling on portable consumer devices such as the iPhone.

This work demonstrates the real-time deployment of a heavily pruned WaveNet-style network for guitar amplifier emulation on a CPU-only iOS implementation. Section~\ref{sec:model} details the training and pruning strategy, and Section~\ref{sec:deploy} the real-time iOS C++ inference engine and its on-device performance. At the demonstration, attendees will play through the on-device model and compare it against the original analog hardware.

\begin{figure}[tb]
\centering
\includegraphics[width=\linewidth]{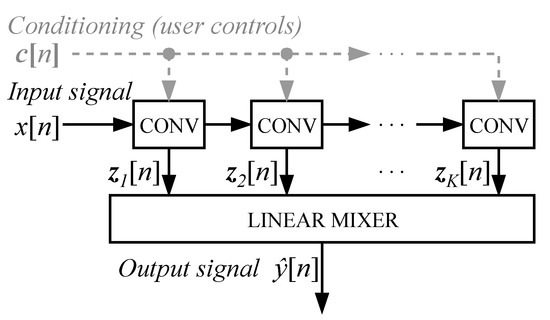}
\caption{WaveNet-style architecture used for neural guitar amplifier modeling, reproduced from \cite{guitar-wavenet}.}
\label{fig:wavenet}
\end{figure}

\section{Model Training and Pruning}
\label{sec:model}

\subsection{Architecture}

We adopted a variant of the original WaveNet architecture~\cite{original-wavenet}, a causal feedforward network composed of stacked dilated convolutional residual blocks (Figure~\ref{fig:wavenet}). The network learned a direct mapping from the raw input guitar waveform $x[n]$ to the corresponding distorted output $\hat{y}[n]$. Dilation expanded the receptive field of each convolution without increasing the number of parameters per layer, allowing the network to capture long temporal dependencies. Each block applied a learned causal FIR filter followed by a nonlinearity, and skip connections aggregated intermediate representations into the final output. We followed the neural-amplifier configuration of Wright et al.~\cite{guitar-wavenet}: a channel dimension $C = 16$, a kernel size of 3, 1500 training epochs, and an 18-layer dilation pattern $d_k = \{1, 2, 4, \ldots, 256, 1, \ldots, 256\}$. Unlike the original architecture, we did not employ gated $\tanh$ activations, instead using standard $\tanh$ nonlinearities to reduce computational overhead in the later C++ deployment. Grounding the network size and topology in this established baseline lets the reported sparsity be interpreted relative to a known reference.

\subsection{Training Objective}
\label{sec:training}

The model was trained by minimizing the mean squared error (MSE),
\begin{equation}
\mathrm{MSE} = \frac{1}{N}\sum_{n=0}^{N-1}\bigl|y_p[n] - \hat{y}_p[n]\bigr|^2,
\label{eq:mse}
\end{equation}
where $y_p[n]$ and $\hat{y}_p[n]$ denote the pre-emphasized target signal and model output, respectively. Pre-emphasis was applied using the first-order infinite impulse response (IIR) high-pass filter
\begin{equation}
H(z) = 1 - 0.95\,z^{-1},
\label{eq:preemph}
\end{equation}
which is commonly used in speech processing and neural amplifier modeling~\cite{Damskagg2018}. Although prior work~\cite{guitar-wavenet} adopted the error-to-signal ratio (ESR),
\begin{equation}
\mathcal{E}_{\mathrm{ESR}} =
\frac{\sum_{n=0}^{N-1} \bigl|y_p[n] - \hat{y}_p[n]\bigr|^2}
     {\sum_{n=0}^{N-1} \bigl|y_p[n]\bigr|^2},
\label{eq:esr}
\end{equation}
as the training loss, our preliminary experiments showed that MSE led to faster and more stable convergence during training. Despite the effectiveness of MSE training, we used ESR as the evaluation metric, since normalizing the error by the signal energy makes it more suitable for comparing models with different output loudness levels.

\subsection{Pruning Strategy}

\begin{figure}[tb]
\centering
\includegraphics[width=\linewidth]{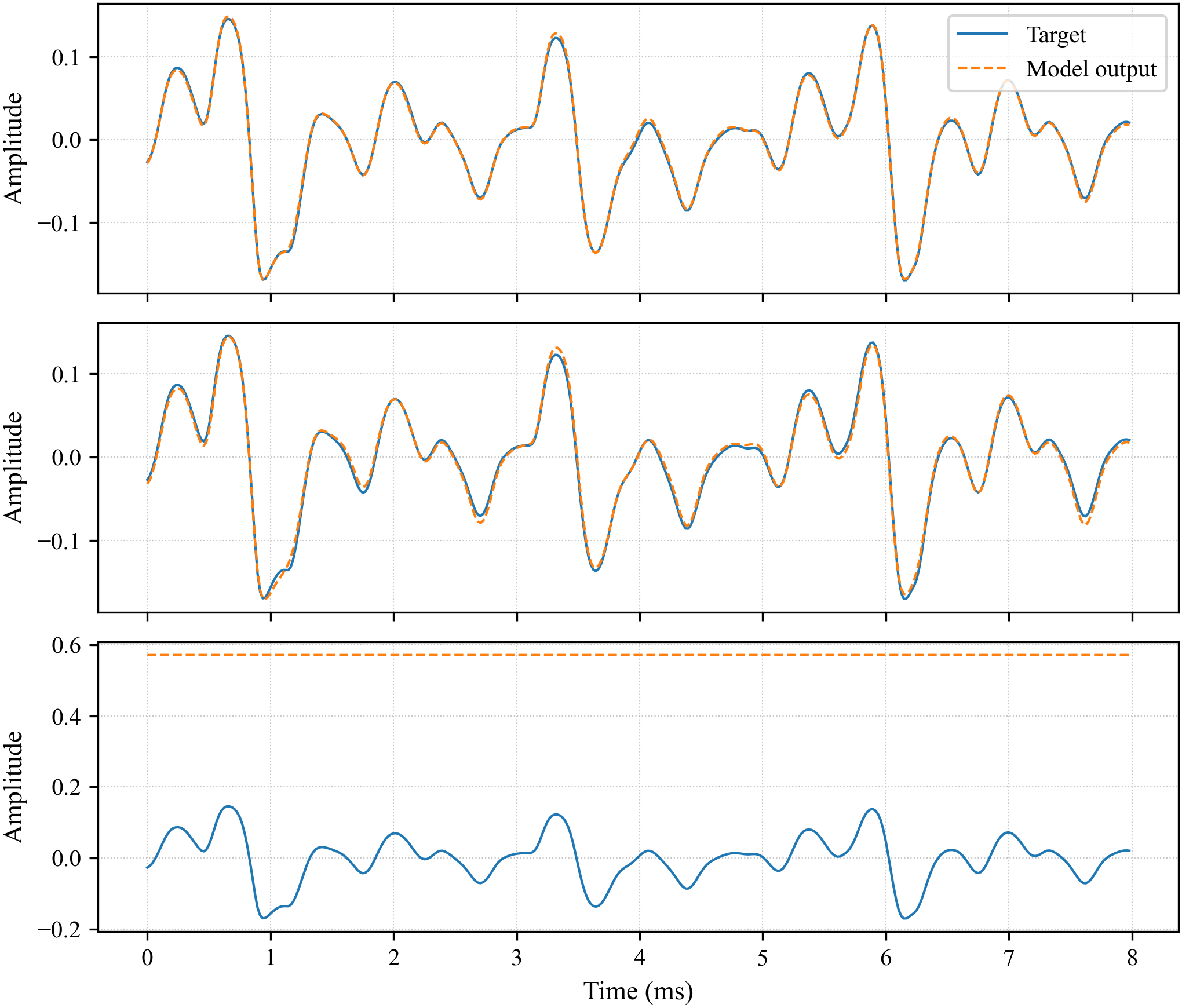}
\caption{Output waveforms for a held-out test segment. \textbf{Top:} unpruned baseline. \textbf{Middle:} 90\% iterative pruning, which tracks the target with only minor deviation. \textbf{Bottom:} 90\% one-shot pruning, which fails to track the target.}
\label{fig:waveform}
\end{figure}

\begin{figure}[tb]
\centering
\includegraphics[width=\linewidth]{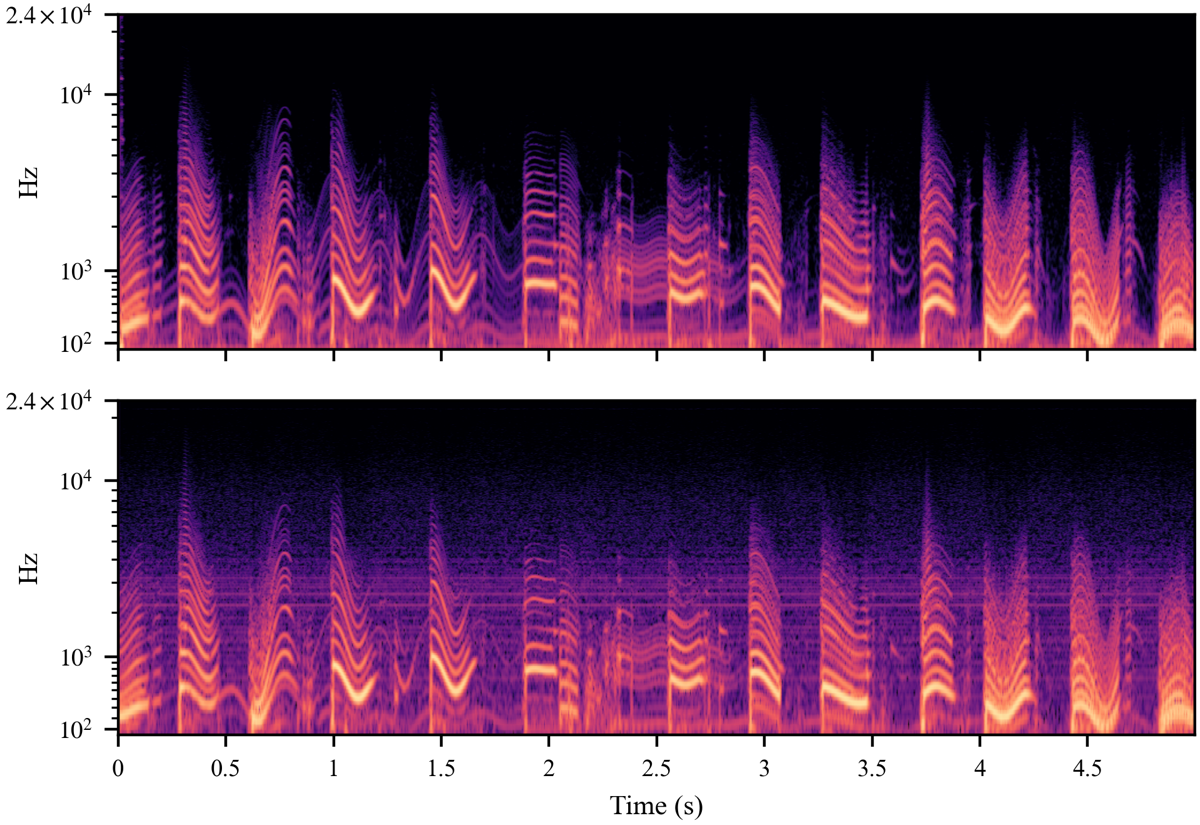}
\caption{Fuzz Face at maximum distortion: 90\% pruned model output (\textbf{top}) versus the recorded target (\textbf{bottom}). The deterministic model omits the input-uncorrelated noise and hum visible in the target.}
\label{fig:fuzzspec}
\end{figure}

To aggressively sparsify the network, we applied iterative local magnitude pruning. Sparsity is defined as the ratio of pruned parameters to the total number of prunable parameters, with biases and other non-prunable parameters excluded. For reference, the 16-channel model contained 21{,}913 total parameters, of which 21{,}152 were prunable. At 90\% sparsity, 19{,}074 prunable weights were removed. 

Each prunable weight tensor $W$ was associated with a binary mask $M$, and the layer used $W \odot M$ in both the forward and backward passes. Rather than pruning once after training (one-shot pruning), the mask was updated every mini-batch according to a sparsity schedule, letting the network adapt to sparsity during training. The sparsity $s$ was ramped from $0$ to the target $s_t$ between epochs $e_{\mathrm{start}}$ and $e_{\mathrm{end}}$ using either a linear or an exponential schedule.

We evaluated all four combinations of local versus global magnitude pruning and linear versus exponential schedules. Fixing $e_{\mathrm{start}} = 10$ and sweeping $e_{\mathrm{end}} \in \{250, 500, 750, 1000, 1250\}$ on a validation split, $e_{\mathrm{end}} = 750$ achieved the lowest validation ESR. Overall, iterative local pruning with an exponential schedule performed best and was used in all subsequent experiments. Iterative pruning preserved accuracy up to high sparsity, whereas one-shot pruning collapsed well before 90\%, as the waveform comparison in Figure~\ref{fig:waveform} shows.

\subsection{Modeling Quality}

We fixed a target of 90\% sparsity, a conservative real-time operating point justified in Section~\ref{sec:runtime}. At this sparsity level, the model reproduced our in-house captures of four guitar distortion sources:
\begin{samepage}
\begin{itemize}
  \item Vox AC15
  \item Fender Deluxe Reverb
  \item Fender Tweed-style amplifier
  \item Dunlop Fuzz Face pedal
\end{itemize}
\end{samepage}
recorded with sE Microphones using a structured three-minute excitation signal~\cite{neuralampmodeler}. All ESR values were below $3.4\times10^{-4}$ with no audible degradation in informal listening. Across captures, the long-reverb settings produced the largest errors, since their tails exceeded the model's receptive field, whereas the directly recorded Fuzz Face attained the lowest ESR despite its strong nonlinearity.

\begin{figure*}[t]
\centering
\includegraphics[width=\textwidth]{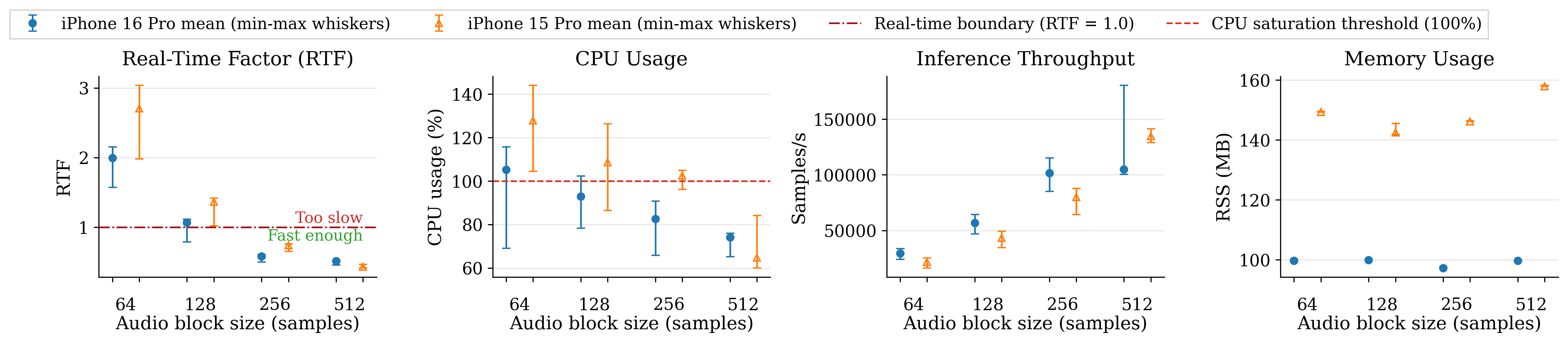}
\caption{Block-size sweep at 90\% sparsity: iPhone~16~Pro versus iPhone~15~Pro, showing RTF, CPU usage, inference throughput, and memory usage. Measurements taken from 60 seconds of audio for each block size.}
\label{fig:blocksize}
\end{figure*}

\begin{figure}[tb]
\centering
\includegraphics[width=0.55\linewidth]{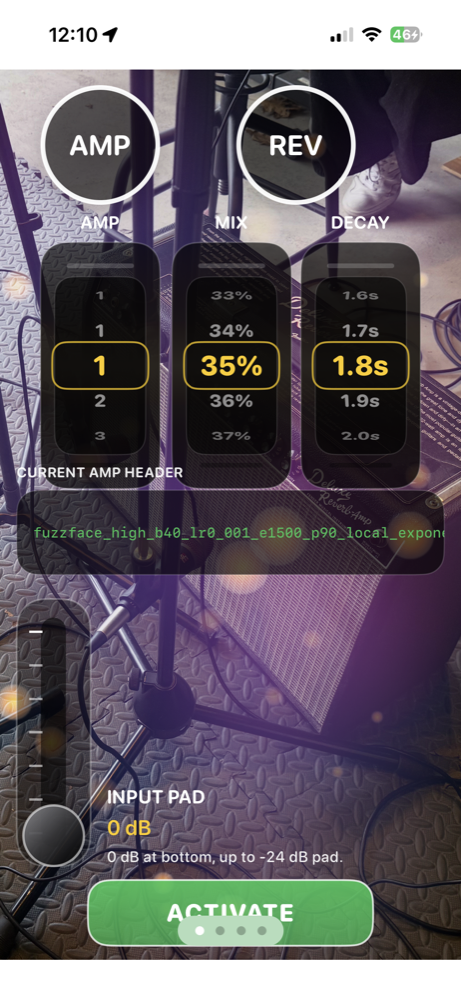}
\caption{iOS application interface: amplifier/pedal selection, input attenuation, and reverb mix and tail-length controls.}
\label{fig:app}
\end{figure}

\begin{figure}[tb]
\centering
\includegraphics[width=\linewidth]{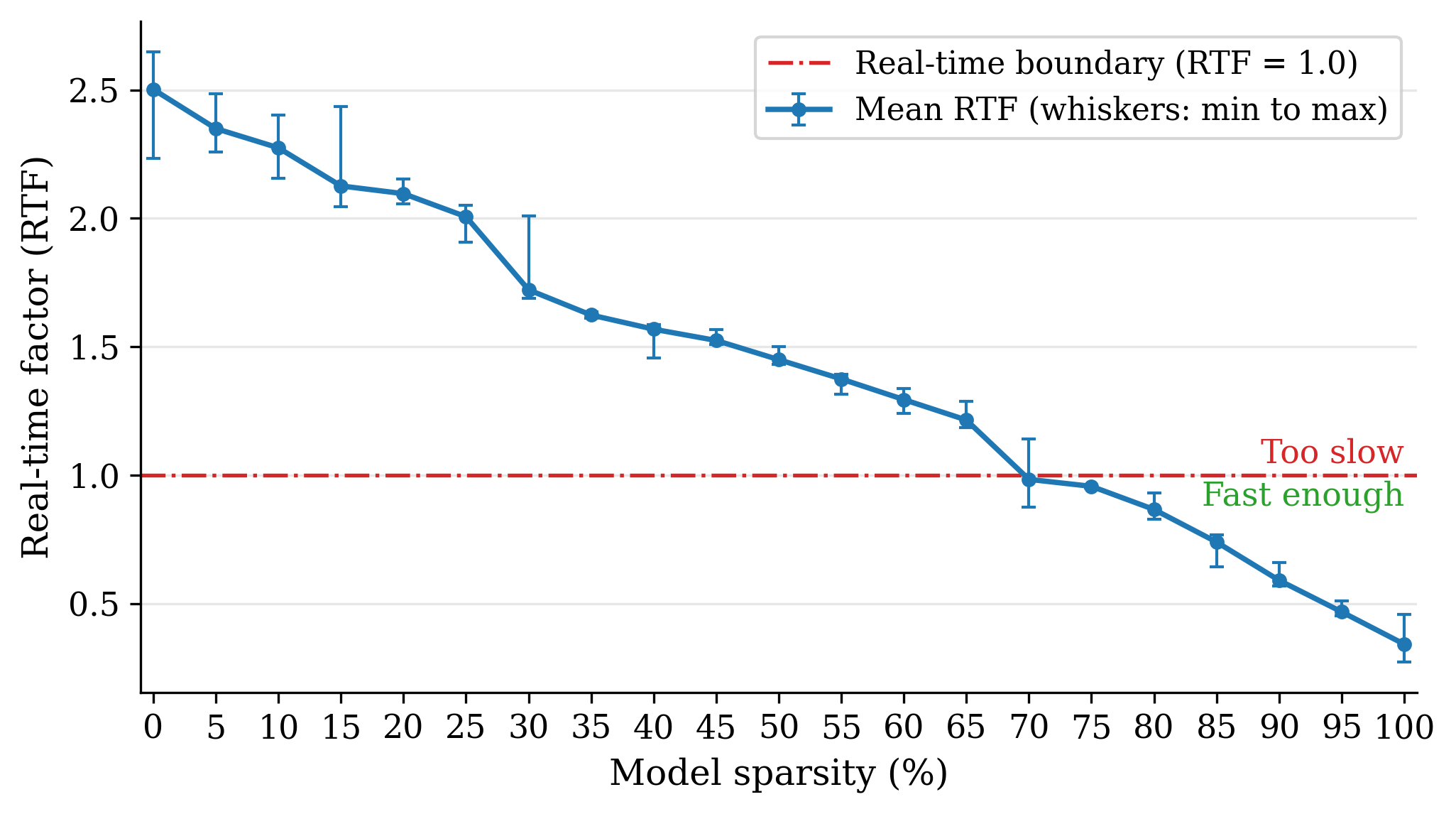}
\caption{Real-time factor versus sparsity on an iPhone~16~Pro (block size 256 samples, \SI{48}{\kilo\hertz}). Points show the mean RTF over 10\,s of audio; whiskers indicate the minimum and maximum. The dense model is intractable, whereas the 90\% pruned model runs with comfortable margin.}
\label{fig:rtf}
\end{figure}

Figure~\ref{fig:waveform} contrasts the output waveforms directly: the 90\% iteratively pruned model tracks the target closely, whereas naive one-shot pruning at the same sparsity fails to track the target waveform. The main residual limitation is that the deterministic model cannot reproduce input-uncorrelated noise and hum, which is most pronounced for the high-gain Fuzz Face as seen in Figure~\ref{fig:fuzzspec}.

\section{Real-Time iOS Deployment}
\label{sec:deploy}

\subsection{Sparse Inference Engine}

Inference runs entirely on the CPU. Although the GPU (Metal) and Neural Engine (Core ML) offer higher throughput and better power efficiency, the per-block dispatch latency of their public interfaces—together with the absence of low-level Neural Engine access—makes them impractical for the small, per-audio-block workloads of real-time inference at a 256-sample block size.

Audio is processed in fixed-size blocks at a \SI{48}{\kilo\hertz} sampling rate. The block size is configurable from 64 to 512 samples, with 256 samples ($\approx$\SI{5.3}{\milli\second}) adopted as the default operating point. Crucially, the engine stores and iterates over \emph{only the nonzero weights}, kept in a compact, cache-friendly layout, rather than masking a dense kernel, so the unstructured sparsity translates directly into compute savings. For this reason we implemented the dilated-convolution stack manually rather than relying on a general-purpose inference library such as RTNeural~\cite{rtneural}, which operates on dense kernels and would not realize the same speedup.

File input can be passed through the C++ inference engine, so deterministic validation against the Python reference is possible. The inference engine inputs and outputs fixed-point audio sample values but converts to floating-point for processing. Against a Python reference using the same exported weights, the C++ engine matched to within the int16/float conversion quantization after engine startup.

\subsection{Interactive Application}

The pruned amplifier was embedded in an interactive iOS application (Figure~\ref{fig:app}) providing a selector for the captured amplifiers/pedals and an input-attenuation control ($0$ to \SI{-24}{\decibel}) for loudness matching across modeled devices. The app also provides a convolutional reverb implemented as partitioned overlap-add convolution with a measured impulse response, exposing wet/dry-mix and tail-length controls. The reverb only accounts for $\approx3\%$ of per-block DSP time.

\subsection{On-Device Performance}
\label{sec:runtime}

We define the real-time factor (RTF) as the ratio of audio-processing time to the duration of audio produced. Thus, RTF\,$<1$ indicates that the engine keeps up with playback. In Figure~\ref{fig:rtf} measured on an iPhone~16~Pro at a 256-sample block size, RTF decreased approximately linearly with sparsity. The dense model sat well above the real-time threshold and was intractable, about 70\% sparsity marked the boundary into real-time operation, and the 90\% point left a comfortable margin (RTF\,$\approx0.6$). This is the central justification for the demonstration, where pruning is what makes an otherwise intractable WaveNet-style model run on the phone.

As seen in Figure~\ref{fig:blocksize}, sweeping the block size across $\{64, 128, 256, 512\}$ samples on both an iPhone 16 Pro and an iPhone 15 Pro revealed that 256 samples was the smallest block that sustained real time on the 16 Pro. Older devices required a larger block or higher sparsity.

\section{Demonstration}
\label{sec:demo}

For the live demonstration attendees will play, or listen to, an electric guitar connected to an iPhone running the neural amp model and listen through headphones. They will switch freely among the captured amplifiers and pedals and adjust the input attenuation and reverb in real time. A Line~6 HX~Stomp will route the signal between the on-phone pruned Fuzz Face model and the physical Fuzz Face pedal, letting attendees compare the on-device emulation directly against the analog hardware the model was trained on.

\section{Conclusion}
\label{sec:conclusion}

This work shows that heavy iterative magnitude pruning (90\% sparsity), paired with a sparse-aware iOS inference engine, lets a WaveNet-style neural guitar amplifier run in real time on a CPU-only iPhone implementation where the dense model cannot, with no perceptible loss in modeling quality, accurate to within int16 quantization error. Future work includes vectorizing the dilated-convolution stack and pointwise $\tanh$ with Apple's Accelerate/vDSP framework, profiling across a wider range of devices, and a formal perceptual evaluation.

\section{Acknowledgments}
This project grew out of the final project for \textit{\href{http://web.stanford.edu/class/ee264/}{EE264W: Digital Signal Processing}} in the Stanford Department of Electrical Engineering, whose real-time iOS audio framework was developed by Fernando Mujica. Many thanks to Professor Fernando Mujica and Ron Schafer for their guidance throughout the project.

\bibliographystyle{IEEEtranDAFx}
\bibliography{DAFx26_tmpl} 

\end{document}